\documentclass{ifacconf}

\usepackage{graphicx}      
\usepackage{subcaption}
\usepackage{subfig}
\usepackage{natbib}        
\usepackage{amsmath}
\usepackage{svg}

\newsavebox{\measurebox}
\begin{document}
\begin{frontmatter}

\title{Optimized Excitation Signal Tailored to Dynamic Operational Characteristics} 

\author[First]{Max Heinz Herkersdorf} \author[First]{Tarek Kösters}  \author[First]{Oliver Nelles} 

\address[First]{Research Group Automatic Control - Mechatronics, University of Siegen, 57076, Siegen, Germany (e-mail: \{max.herkersdorf, tarek.koesters, oliver.nelles\}@uni.siegen.de).}

\begin{abstract}                
The identification of nonlinear dynamic systems through data-driven methods is significantly impacted by the excitation signal used to generate training data. The Incremental Dynamic Space-Filling Design (IDS-FID) strategy introduces a Design of Experiment (DoE) technique aimed at achieving a space-filling distribution within the input space of the nonlinear approximator used in external dynamics modeling. Simultaneously, the approach enables control over the excited frequency spectrum. Hence, the IDS-FID algorithm is able to influence the dynamics of the generated excitation. Application of this algorithm on artificial test data reveals that tailoring the dynamics of an excitation signal to match the expected frequency range of process operation results in enhanced model accuracy. Overall, the IDS-FID strategy proves to be highly competitive, surpassing the performance of state-of-the-art DoE techniques.
\end{abstract}


\end{frontmatter}

\section{Introduction}
Contemporary nonlinear system identification applications leverage powerful machine learning techniques to a great extent. The performance of these data-driven approaches is highly dependent on the quality of the data utilized for model training. Hence, apart from the decision on model structure and estimation of the parameters, the design of suitable excitation / input signals is of crucial importance. Nonetheless, especially for nonlinear dynamic processes, this is still an under-researched field.

The decisive criterion in static Design of Experiments (DoE) is the data distribution of the input space. Without prior process knowledge, it is common practice to aim for a space-filling distribution in order to collect data with equal density in every operating region. Popular approaches are Sobol sequences \citep{sobol1967distribution} and optimized Latin hypercube designs \citep{johnson1990minimax}.

Design of excitation signals for nonlinear system identification demands advanced requirements: (i)  To encompass the full extent of nonlinear characteristics, it is necessary to collect data across the entire operational area. This endeavor is complicated by the fact that dynamic models require delayed process inputs and possibly outputs, which cannot be manipulated independently. (ii) To obtain insights into the dynamic behavior of the process, it is essential to stimulate the pertinent frequency range. Consequently, the complete trajectory of the input signal holds significance. \\
Typical state-of-the-art excitation signals for the identification of nonlinear systems focus primarily on only one of the aforementioned properties, e.g., amplitude pseudorandom binary signals (APRBS)  \citep{nelles1995identification} target the relevant amplitudes, while chirp \citep{baumann2008excitation} and multisine (MS) \citep{pintelon2012system} signals cover the frequency range of interest.

In \cite{heinz2017iterative}, the authors introduce an excitation signal design strategy for nonlinear dynamic processes referred to as \textit{OptiMized Nonlinear InPUt Signal} (OMNIPUS). The method aims to optimize the data point distribution within the input space of the nonlinear approximator used in external dynamics modeling towards a space-filling distribution. Without prior process knowledge, this seems reasonable, since data is collected across the operational area. This allows the nonlinearity to be entirely approximated and thus extrapolation to be limited. 

In this work, an extension of the described OMNIPUS strategy is proposed: the new method also pursues a space-filling design in the input space of the nonlinear approximator, but additionally allows the manipulation of the excited frequency spectrum. This capability enables customization of the excitation to match the anticipated frequency range of process operation, in the following also referred to as dynamic operational charactersitics, facilitating an improvement in model performance. The novel approach is called \textit{Incremental Dynamic Space-Filling Design} (IDS-FID) and is characterized by simple operation in real-world applications, demanding only little process knowledge.

The contribution begins by elucidating the foundational proceeding of IDS-FID in Sect.$\ $\ref{SecIDSFID}, with a specific emphasis on how it tackles the space-filling criterion and the control over the excited frequency spectrum. Section \ref{SecEvaluation} explains the evaluation methodology under which the new method was investigated, followed by the presentation of the findings. Finally, the main results are summarized and outlook on further research is provided in Sect.$\ $\ref{SecConclusion}.

\section{Incremental Dynamic Space-Filling Design (IDS-FID)} \label{SecIDSFID}
The core idea of the suggested DoE technique centers around the iterative concatenation of optimal input sequences (optimal input signal parts). In this context, optimal means maximizing a distance-based quality function that relies on a rough process model and is inspired by the maximin criterion proposed in \cite{johnson1990minimax}. The foundational optimization procedure is analogous to the OMNIPUS approach and is succinctly outlined in Sect.$\ $\ref{SubSecOptimizationIDSFID}. Subsequently, in Sect.$\ $\ref{SubSecQualityFunction}, a novel quality function is introduced, which is purposefully crafted to satisfy the space-filling criterion in the input space of the nonlinear approximator while simultaneously offering control over the influenced frequency spectrum.
\subsection{Optimization Strategy} \label{SubSecOptimizationIDSFID}
\begin{figure}[h]
	\begin{center}
		\includegraphics[width=8.4cm]{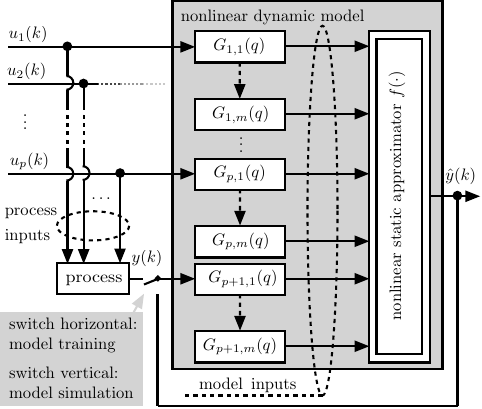}    
		\caption{ External dynamics approach for $p$ inputs, one output and a common dynamic order $m$ for simplicity.} 
		\label{fig:ExtDyn}
	\end{center}
\end{figure}

The primary objective in modeling a nonlinear system is to discover a function $\hat{y} = f(\underline{x})$ that approximates the process output $y$ based on the inputs $\underline{x}$. In a data-driven modeling approach, the function $f(\underline{x})$ is heavily dependent on the data utilized for model training. \\
A prevalent dynamic modeling approach, as illustrated in Fig.$\ $\ref{fig:ExtDyn}, employs external dynamics alongside a nonlinear static approximator. The input space of the nonlinear static approximator, hereafter referred to as the regressor space, is encompassed by filtered inputs and filtered outputs. In this paper, the \textit{Nonlinear AutoRegressive with eXogenous input} (NARX) structure is adopted, resulting in the external filters being simple delay elements $G_{i,j}(q) = q^{-1}$. 

\textbf{Regressor Space of Dynamic Systems} \hspace{10pt} Because a fundamental characteristic of the IDS-FID strategy is rooted in calculating distances within the regressor space, it is essential to note that process or model outputs are not accessible until after measurement. Therefore, a roughly estimated model, referred to as the proxy model, is employed for signal optimization. In the most basic scenario, a linear first order transfer function is assumed as a proxy model for each input. This simplified model facilitates the computation of the proxy output $\tilde{y}$ for any specified input, enabling the establishment of a data point distribution in the proxy regressor space, with $\tilde{y}$ substituting the unknown $y$.  By default, the gains are set to 1, ensuring that both $u$- and $\tilde{y}$-directions carry equal importance in distance calculation. Consequently, minimal process knowledge is required (in the simplest form), with only time constants from each input to the output have to be assumed. \cite{heinz2017iterative} demonstrate that the disparities between the data point distribution in the actual regressor space (with process output) and the proxy regressor space (with linear model output) are minimal when both have similar time constants. \\
A single sample at time step $k$ of an excitation signal with $p$ inputs can be written as 
\begin{align} \label{eq:inuptp}
\underline{u}^{T}(k) = [u_{1}(k), u_{2}(k), \ldots , u_{p}(k)] \ .
\end{align}
With a dynamic order of $m$, the proxy input therefore equals
\begin{align} \label{eq:proxyInput}
	\begin{split}
\tilde{\underline{x}}^{T}(k) =   &  \big[	u_{1}(k-1), \ldots, u_{1}(k-m), u_{2}(k-1), \ldots, \\
 & u_{2}(k-m), \ldots, 	u_{p}(k-1),  \ldots, u_{p}(k-m), \\
  & \tilde{y}(k-1), \ldots, \tilde{y}(k-m)	\big] \ .
 	\end{split}
\end{align}
 Consequently, a signal's proxy input space distribution can be depicted by the following matrix
\begin{align} \label{eq:proxySpace}
	\begin{split}
 \underline{\tilde{X}}= [
		\tilde{\underline{x}}(1), \tilde{\underline{x}}(2), \ldots, \tilde{\underline{x}}(N) ] \ . 
		 	\end{split}
	\end{align}
	
\begin{figure}[h]
	\centering
	\begin{subfigure}[t]{0.2\textwidth}
		\centering
		\includegraphics[width=4cm, trim=6cm 12.2cm 6cm 12.2cm]{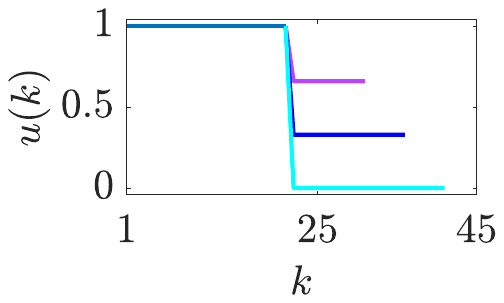}
		\caption{Already optimized signal and three new amplitude levels in combination with their optimal length.}
		\label{fig:SequenceSelectionSignal}
	\end{subfigure}
	\hfill 
		\begin{subfigure}[t]{0.2\textwidth}
		\centering
		\includegraphics[width=4cm, trim=6cm 10cm 6cm 10cm]{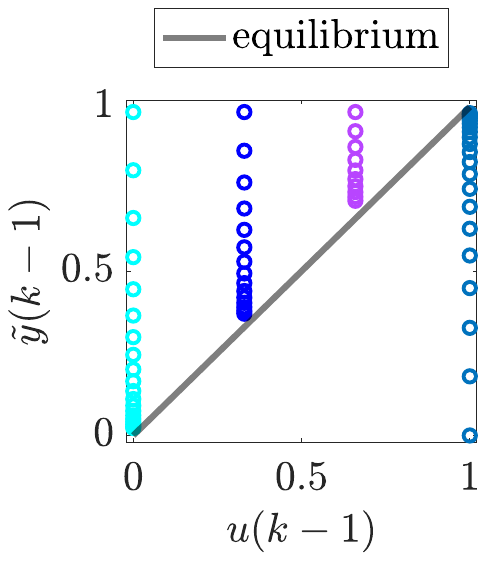}
		\caption{Data point distribution of the already optimized signal and the possible new sequences in proxy regressor space.}
		\label{fig:SequenceSelectionProxy}
	\end{subfigure}
		\caption{ Illustrative example of a sequence selection with one input. The proxy model presented here equals the one employed in the evaluation.}
	\label{fig:SequenceSelection}
	\end{figure}

\textbf{Concatenation of Optimal Sequences} \hspace{10pt} 
Optimizing an excitation signal with $N$ time steps for $p$ inputs leads to an optimization problem of size $N \times p$. In the case of multivariate processes and substantial values of $N$, the quest for the global optimum, or even favorable local optima, becomes infeasible due to computational complexity. Hence, a robust approach is introduced,  which breaks down the extensive global optimization problem into a series of smaller, low-dimensional optimizations, increasing the likelihood of achieving a favorable local optimum while significantly decreasing the computational complexity. 
\begin{figure} [h]
	\begin{center}
		\includegraphics[width=10.0cm, trim=1cm 21cm 1cm 6cm]{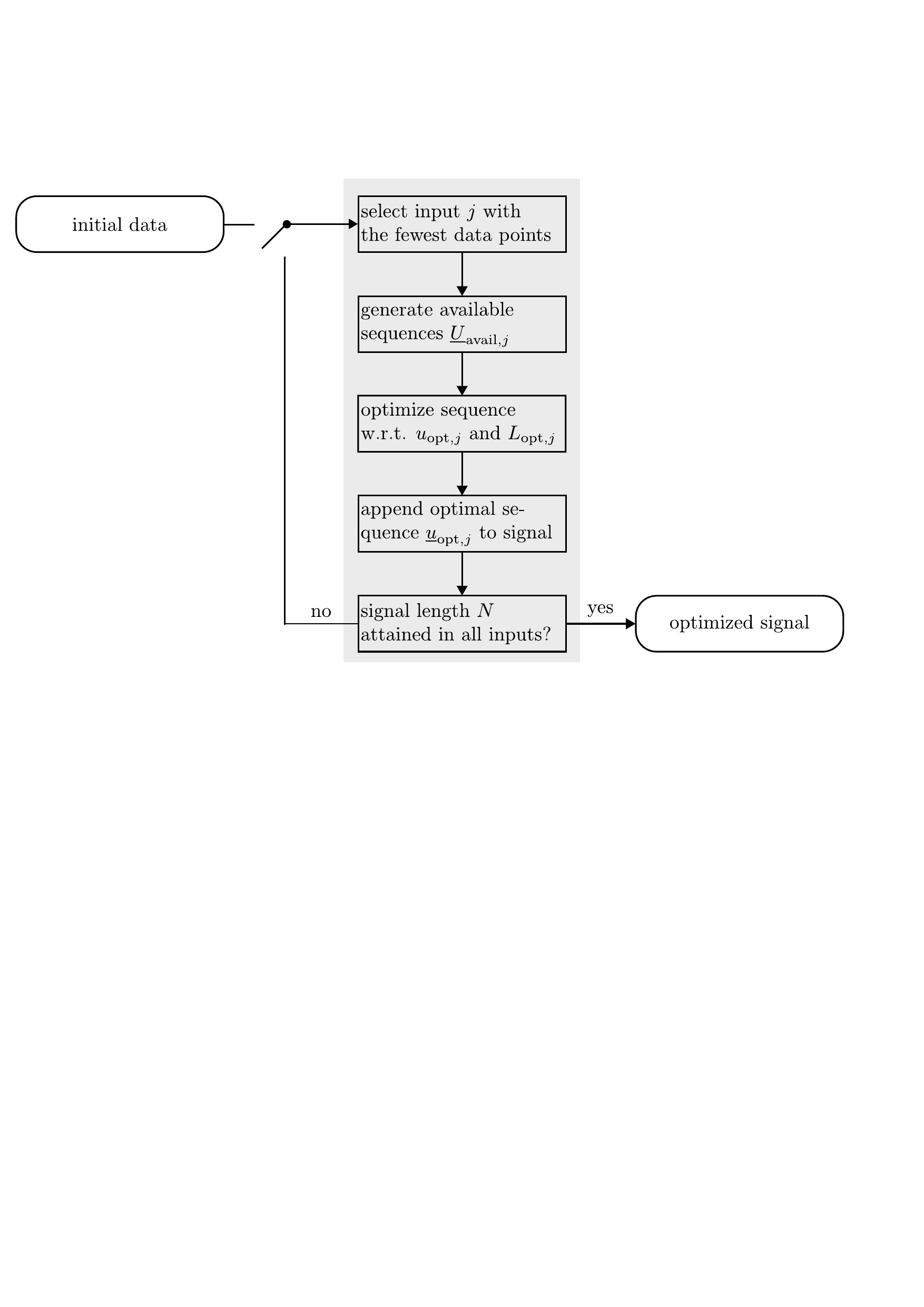}    
		\caption{Flow chart of the iterative optimization.} 
		\label{fig:Optimization}
	\end{center}
\end{figure}
 \\ 
An excitation signal created with the IDS-FID algorithm consists of multiple optimal sequences. As illustrated in Fig.$\ $\ref{fig:Optimization},  the excitation signal generation is executed iteratively, with each iteration (gray shaded box) producing an optimal sequence exclusively for a single input. Newly generated signal parts are therefore appended to the existing data input by input, potentially resulting in different lengths in the individual inputs during the process of signal generation. \\
Thus, at the commencement of each sequence generation, the input $j$ with the currently fewest data points is chosen. If multiple inputs share the same length, one of them is randomly selected. During the entire iteration, the other inputs remain constant.\\
Afterwards the set of sequences available in this iteration $\underline{U}_{\mathrm{avail},j}$ is generated for the selected input. To further simplify the optimization process, the sequences are defined as piece-wise constant, resulting in an APRBS-shaped signal. Hence, $\underline{U}_{\mathrm{avail},j}$ results from each possible combination of available amplitude values $\underline{u}_{\mathrm{avail}, j}$ and available sequence lengths $\underline{L}_{\mathrm{avail}, j}$. By default the number of possible time steps per sequence is configured as $\underline{L}_{\mathrm{avail}, j}  = [1, 2, \ldots, L_{\mathrm{max},j}]$ with $L_{\mathrm{max}} = 3{T_{j}}/{T_{0}}$. Here, $T_{j}$ represents the time constant of the transfer function for input $j$ to the output and $T_{0}$ denotes the sampling time. The relatively long maximum sequence length $L_{\mathrm{max},j}$  is advantageous for effectively identifying the steady state. As mentioned in \cite{morris1995exploratory}, for higher-dimensional input spaces, optimizations based on the maximin criterion tend to produce designs with a majority of data points clustered at the boundaries and corners of the input space. This phenomenon is a manifestation of the curse of dimensionality, making it exceedingly challenging to identify the nonlinearity. To overcome this undesired behavior, in  \cite{heinz2018excitation} a \textit{Latin hypercube} inspired approach is proposed, which is applied here as well and exhibits two characteristics:
(i) The amplitude values of the optimized signal are restricted to operate on predefined, typically equidistantly chosen levels. (ii) After a specific amplitude level has been used, revisiting it is blocked until all other levels have been employed. Consequently, by specifying a number of visitable amplitude values $M_{j}$, the attainable levels can be established. In the instance of an amplitude range $0 \leq u_{j} \leq 1$ and in the absence of any levels being blocked, this results in
\begin{align} \label{eq:availableAmplitudeLevels}
	\begin{split}
		& \underline{u}_{\mathrm{avail}, j}^{T}  = [0, \Delta u_{j}, 2 \Delta u_{j}, \ldots, (M_{j}-2)\Delta u_{j},1] \\
		& \mathrm{with} \ \Delta u_{j}=\frac{1}{M_{j}-1} \ .
	\end{split}
\end{align}
Without user interaction, the resolution is determined by employing an average sequence length of $ \overline{L}_{j} = {T_{j}} / {T_{0}}$, resulting in $M_{j} = \lceil {N}/{\overline{L}_{j}} \rceil $ with $\lceil \cdot \rceil$ representing the ceiling function. \\
Therefore, solely the new amplitude value $ u_{\mathrm{opt},j}$ and the new sequence length $L_{\mathrm{opt},j}$ have to be optimized. These are selected with respect to the maximum of the quality function $J$ according to
\begin{align} \label{eq:optimalSeq}
	\begin{split}
		& \underline{u}_{ \mathrm{opt},j } = \underset{u_{\mathrm{new},j} ,  L_{\mathrm{new},j} }{ \mathrm{argmax}}  J( \underline{U}_{\mathrm{avail},j} )    \\
		& \mathrm{subject \ to\ } u_{\mathrm{opt},j} \in \underline{u}_{\mathrm{avail}, j}\\	 
		& \hspace{47pt} L_{\mathrm{opt},j} \in \underline{L}_{\mathrm{avail}, j}  \  .
	\end{split}
\end{align}
The computation of $J$ takes place in the proxy regressor space and is elaborated further in Sect.$\ $\ref{SubSecQualityFunction}. An illustrative example with one input featuring an already optimized signal part and potential new sequences up to their optimal length, along with associated distributions in the proxy regressor space, is provided in Fig.$\  $\ref{fig:SequenceSelection}.
\\ The emerging optimal sequence $\underline{u}_{\mathrm{opt},j}$ is subsequently appended to the (already optimized) data points of input $j$.\\
This procedure is reiterated until the desired excitation signal length $N$ is achieved for all inputs. \\
Without initial data, the IDS-FID generation starts with the steady state of the process at the start of measurements. However, it can also be used to complete initial data in an optimal manner. \\
Since the iterative optimization of sequences and the exploitation of a proxy regressor space closely resemble the OMNIPUS strategy, the advantages outlined for industrial applications in \cite{kosters2022optimization} and for multidimensional input spaces in \cite{heinz2018excitation} are equally transferable to the IDS-FID approach.
\subsection{Quality Function} \label{SubSecQualityFunction}
The novel quality function can be decomposed into two distinct parts and tuned with the hyperparameter $\lambda$
\begin{align} \label{eq:qualityFunction}
	J = J_{1} -\lambda \cdot J_{2} \ .
\end{align}
To account for a space-filling distribution the first part of the quality function is defined as follows:
\begin{align} \label{eq:rewardTerm}
	J_{1} = \sum^{N+L}_{k=N+1} d_{\mathrm{NN}}\bigl( \tilde{\underline{X}}, \underline{\tilde{x}}(k) \bigr) \ .
\end{align}
The function $d_{\mathrm{NN}}\bigl( \tilde{\underline{X}}, \underline{\tilde{x}}(k) \bigr)$ computes the nearest neighbor distance, by default exploiting Euclidean metric, between each existing point in $\tilde{\underline{X}}$ and a data point of the new sequence $\underline{\tilde{x}}(k)$ within the proxy regresspr space. Hence, $J_{1}$ attains its maximum value for the sequence that contains the data points with the greatest cumulative distance to the previously optimized signal. In other words, it reaches its maximum for the sequence that fills the largest gap in the proxy regressor space. \\
The second part of the quality function, facilitates the manipulation of the excited spectrum, specifically aiming to gather more information about the dynamic behavior. Since $J_{1}$ exclusively considers sequences of length $L_{\mathrm{max},j}$ as optimal, this is done by introducing a length-dependent penalty that favors shorter, more dynamic sequences according to
\begin{align} \label{eq:penaltyTerm}
	J_{2} = F \cdot (L^{n})_{\mathrm{scaled}} \ .
\end{align}
The sequence length is raised to the power $n$ and then scaled so that its range is in the intervall $[0, 1]$. The exponentiation grants control over the distribution of selected sequence lengths: For $n=1$, the entire range of $\underline{L}_{\mathrm{avail}, j}$ is exploited, whereas an increasing value of $n$ results in a reduced occurrence of short sequences.
\begin{figure}[h]
	\centering
	\begin{minipage}[t]{0.2\textwidth}
	\centering
	\includegraphics[width=4cm, trim=6cm 11.2cm 6cm 11.2cm]{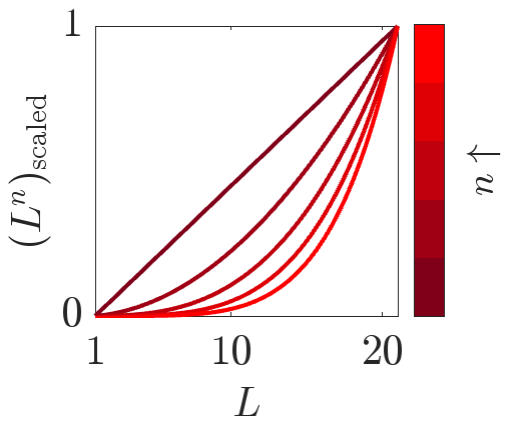}
	\caption{Influence of power $n$ on the shape of $J_{2}(L)$.}
	\label{fig:nInfluence}
\end{minipage}
	\hfill 
		\begin{minipage}[t]{0.2\textwidth}
		\centering
		\includegraphics[width=4cm, trim=6cm 11.2cm 6cm 11.2cm]{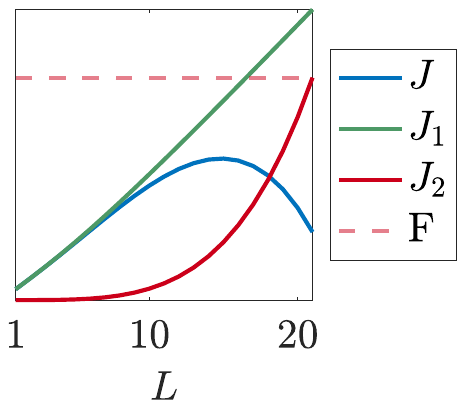}
		\caption{Course of $J(L)$ and its components for an exemplary $u$-value and $n=4$.}
		\label{fig:Jcourse}
	\end{minipage}
\end{figure}
This is caused by the fact that a rising $n$ leads to decreasing penalties for small lengths, as is evident in Fig.$\ $\ref{fig:nInfluence}. The avoidance of (too) short sequences is often reasonable in industrial applications to prevent actuator damage. An exemplary course of $J$ and its components over $L$ for a single $u$-value and the by default chosen $n=4$ can be seen in Fig.$\ $\ref{fig:Jcourse}. Remember that this course of $J(L)$ is computed for all values in $\underline{u}_{\mathrm{avail}, j}^{T}$, and the overall maximum value is considered as optimal, as indicated in Eq.$\ $(\ref{eq:optimalSeq}). \\
However, the penalty term Eq.$\ $(\ref{eq:penaltyTerm}) is designed to consider not only the length of the sequence but also whether the process is already settling (steady-state) or still in transition (transient behavior). A scaling factor $F$ is thus introduced, which assesses the point spacing within the sequence
\begin{align} \label{eq:FTerm}
	F = \biggl( \frac{1}{L_{\mathrm{max}}-1}  \sum^{N+L_{\mathrm{max},j}}_{k=N+2} d \bigl( \underline{ \tilde{x}}(k), \underline{\tilde{x}} (k-1) \bigr) \biggr)^{-1} \ 
\end{align}
with $d (\cdot)$ beeing the distance function. Therefore, $F$ represents the inverse average distance between new data points of a sequence up to $L_{\text{max}}$. According to Eq.$\ $(\ref{eq:FTerm}), $F$ remains constant for all potential lengths at a given $u$-value and serves as a dynamic or activity assessment for these sequences. To be more detailed, the course of $J_{2}$ across $L$ tends to be higher on average for sequences with low dynamics and smaller for those displaying high dynamics. This results in the following outcomes: (i) Sequences characterized by low dynamics are comparatively highly penalized and allocated relatively short lengths. Conversely, sequences with high activity and thus lower penalties are assigned longer lengths. Thus, the placement of data points near the static equilibrium, offering only limited dynamic process insights, is alleviated. Instead, dynamic segments of the sequences are prioritized. (ii) Preference is given to selecting sequences with high activity (little penalty values), thereby achieving the desired information gain about the dynamic behavior. 
\begin{figure}[h]
	\centering
	\begin{subfigure}[t]{0.2\textwidth}
		\centering
		\includegraphics[width=4cm, trim=6cm 12.2cm 6cm 12.2cm]{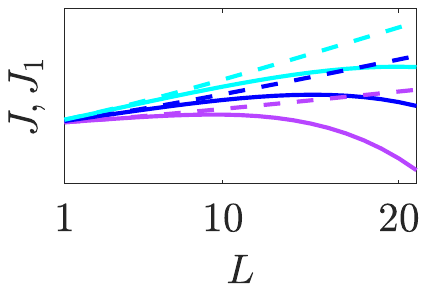}
		\caption{Courses of $J(L)$ and with dashed lines those of $J_{1}(L)$.}
		\label{SequenceSelectionJ}
	\end{subfigure}
	\hfill
		\begin{subfigure}[t]{0.2\textwidth}
		\centering
		\includegraphics[width=4cm, trim=6cm 12.2cm 6cm 12.2cm]{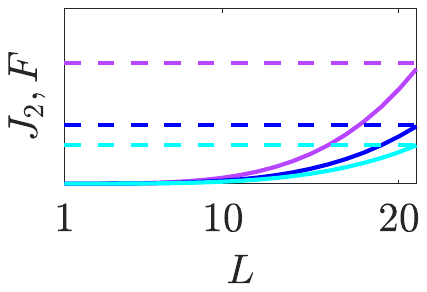}
		\caption{Courses of $J_{2}(L)$ and with dashed lines the associated $F$-values.}
		\label{SequenceSelectionJ12}
	\end{subfigure}
	\caption{Illustration of the quality function and its components for the possible new sequences in Fig.$\ $\ref{fig:SequenceSelection}.}
	\label{fig:SequenceSelectionJ}
\end{figure}
Figure \ref{fig:SequenceSelectionJ} depicts the progression of the quality functions and their components for the example sequences shown in Fig.$\ $\ref{fig:SequenceSelection}.\\
It is noteworthy that the values of $J_{1}$ generally decrease as the signal progresses and thus the proxy input space becomes fuller, whereas $J_{2}$ remains unaffected by this. Hence, as the space-filling criterion is increasingly met, greater emphasis is placed on selecting more dynamic sequences in the optimization. In this context, the hyperparameter $\lambda$ determines the rate at which the emphasis in sequence selection transitions from the space-filling criterion to the selection of dynamic sequences. To put it more practically, the $\lambda$-value determines at what overall signal length shorter and more dynamic sequences are favored.

\section{Evaluation}  \label{SecEvaluation}
In general, good excitation signals result in datasets that carry valuable information, contributing to the development of high-grade models. Consequently, in this paper, model performance is utilized as a metric to gauge the excitation quality in comparison to other state-of-the-art DoE methods. For a more profound analysis, additional consideration is given to the signal trajectory, the spectrum it excites, and the distribution within the input space of the nonlinear approximator. The evaluation methodology employed in this work is detailed in Sect.$\ $\ref{SubSecEvaluationMeth}. Following that, Sect.$\ $\ref{SubSecFindings} presents and interprets the discovered outcomes.

\subsection{Evaluation Methodology} \label{SubSecEvaluationMeth}
A nonlinear dynamic Hammerstein process according to
\begin{align} 	\label{eq:Hammerstein}
	\begin{split}	
		& y(k) = 0.2f(u(k-1))+0.8y(k-1), \\
		& f(x) =  \frac{\mathrm{atan}(8x-4)+\mathrm{atan}(4)}{2\mathrm{atan}(4)} 
	\end{split}
\end{align}
serves as artifical test process. The proxy model is thus a first-order single-input single-output (SISO) system and exhibits a linear transfer function (cf.,$\ $Fig.$\ $\ref{fig:SequenceSelectionProxy}). Given that the focus of this contribution is to highlight the effects of a manipulated spectrum on excitation signal quality, it is sufficient to conduct all evaluations with only a single input. 
\begin{table}[hb]
	\begin{center}
		\caption{Properties of Training Data}\label{tb:trainData}
		\begin{tabular}{c c}
			Notation & Signal   \\ \hline
			$\underline{u}_{\lambda_{1}}$& IDS-FID signal with $\lambda = 0$ \\ \hline
			$\underline{u}_{\lambda_{2}}$& IDS-FID signal with $\lambda = 0.02$  \\ \hline
			$\underline{u}_{\lambda_{3}}$& IDS-FID signal with $\lambda = 0.5$  \\ \hline
			$\underline{u}_{\lambda_{4}}$& IDS-FID signal with $\lambda = 2$  \\ \hline
			$\underline{u}_{\mathrm{OMNIPUS}}$ & OMNIPUS\\ \hline
			$\underline{U}_{\mathrm{APRBS}}$& APRBS with $T_{H} = 5s$  \\ \hline
			$\underline{u}_{\mathrm{MS}}$& Multisine exciting $[0, 0.5]$ Hz\\ \hline
		\end{tabular}
	\end{center}
\end{table}
\\ For the modeling, local model networks (LMNs) in NARX structure, employing the local linear model tree (LOLIMOT) algorithm \citep{nelles2020nonlinear}, are trained with the excitation signals outlined in Tab.$\ $\ref*{tb:trainData}. To accommodate realistic conditions, white noise distributed $\mathcal{N}(\mu = 0, \ \sigma^{2} = 0.01)$ is added to the process output. The evaluation procedure involves generating 100 training data sets and accordingly training 100 LMNs for each excitation. Concerning the deterministic excitation signals\footnote{I.e., $\underline{u}_{\lambda_{1}}$, $\underline{u}_{\lambda_{2}}$, 	$\underline{u}_{\lambda_{3}}$, $\underline{u}_{\mathrm{OMNIPUS}}$, and $\underline{u}_{\mathrm{MS}}$.}, the 100 training data sets per excitation differ only in their noise realizations.
In contrast, owing to its randomness, 100 different realizations of an APRBS with a minimum holding time $T_{H}=5s$ are stored in $\underline{U}_{\mathrm{APRBS}}$ and each of them is assigned a noise realization. All training data sets consist of $N=300$ samples.  \begin{table}[h]
	\begin{center}
		\caption{Properties of Test Data}\label{tb:testData}
		\begin{tabular}{c c}
			Notation & Signal   \\ \hline
			$\underline{u}_{\mathrm{Ramp,t}}$& Ramp signal in $[0,1]$ \\ \hline
			$\underline{U}_{\mathrm{APRBS,t}}$& APRBS with $T_{H} = 1s$  \\ \hline
			$\underline{u}_{\mathrm{MS_{1}, t}}$&Multisine exciting $[0.05, 0.1]$ Hz  \\ \hline
			$\underline{u}_{\mathrm{MS_{2}, t}}$& Multisine exciting $[0.25, 0.3]$ Hz \\ \hline
		\end{tabular}
	\end{center}
\end{table}
\\ To highlight how the excitation affects the model accuracy in different dynamic operational areas, all trained LMNs are evaluated on distinct test data, see Tab.$\ $\ref*{tb:testData}. Here, $\underline{u}_{\mathrm{Ramp,t}}$ is used to asses the approximation of the steady state, while 	$\underline{U}_{\mathrm{APRBS,t}}$ tests step-shaped data with short holding times. The sinusoidal $\underline{u}_{\mathrm{MS_{1}, t}}$ and $\underline{u}_{\mathrm{MS_{2}, t}}$ illustrate the performance of the excitation signals on various operating frequency ranges in detail. For each trained LMN, the model output $\tilde{y}$ is simulated on all test data and the Root Mean Squared Error (RMSE) is subsequently calculated in comparison to the real process output $y$ to assess performance. The test data set size is set to $N_{t} = 500$. \\
Apart from tuning $\lambda$, the IDS-FID algorithm is executed in its described base settings. The operating range for all excitation signals is set to $[0,1]$, and a sampling time $T_{0} = 1s$ is chosen. 

\subsection{Findings} \label{SubSecFindings}
\begin{figure}[h]
	\centering
	\begin{subfigure}{0.4\textwidth}
		\includegraphics[width=8.4cm, trim=3cm 11.2cm 0.5cm 11.2cm]{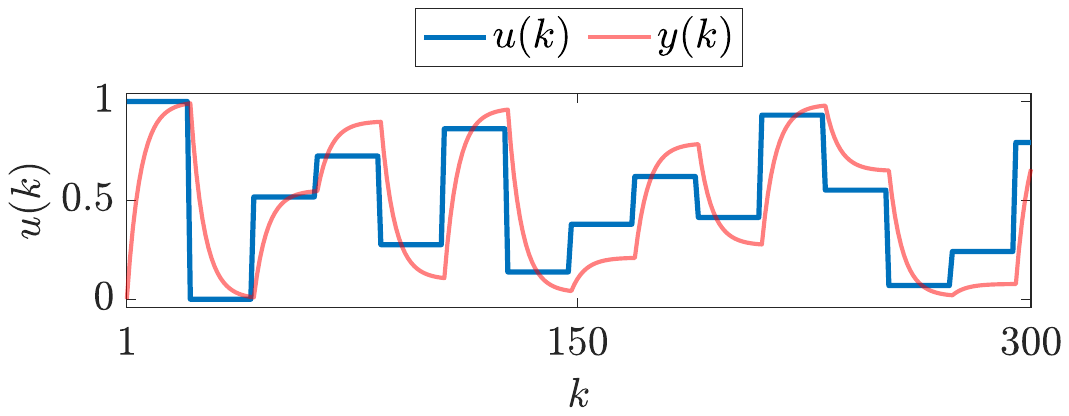}
		\caption{}
		\label{subfig:signalLambda0}
	\end{subfigure}
	\hfill
	\begin{subfigure}{0.4\textwidth}
		\includegraphics[width=8.4cm, trim=3cm 11.2cm 0.5cm 11.2cm]{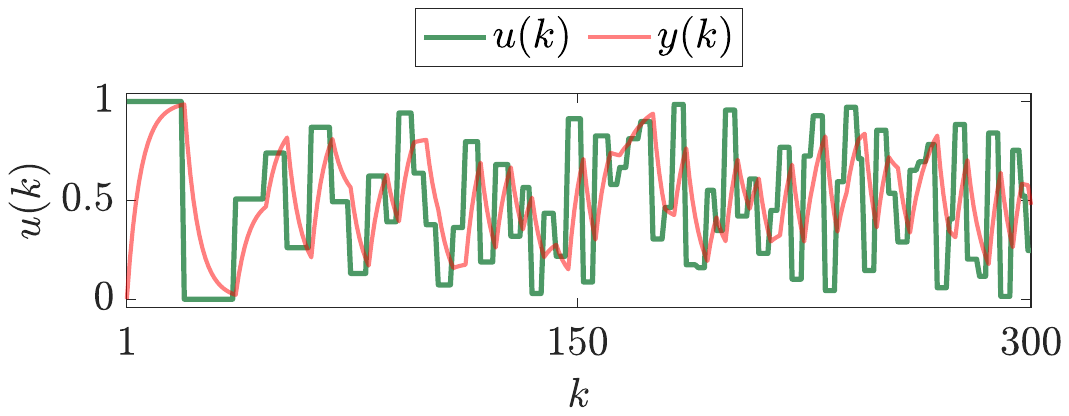}
		\caption{}
		\label{subfig:signalLambda05}
	\end{subfigure}
	\caption{Trajectory of $\underline{u}_{\lambda_{1}}$ in (a) and $\underline{u}_{\lambda_{3}}$ in (b) in combination with the corresponding process output of the Hammerstein system.}
	\label{fig:Signal}
\end{figure}

\begin{figure}[h]
	\centering
	\hfill
		\includegraphics[width=8.4cm, trim=2cm 12cm 0.5cm 12cm]{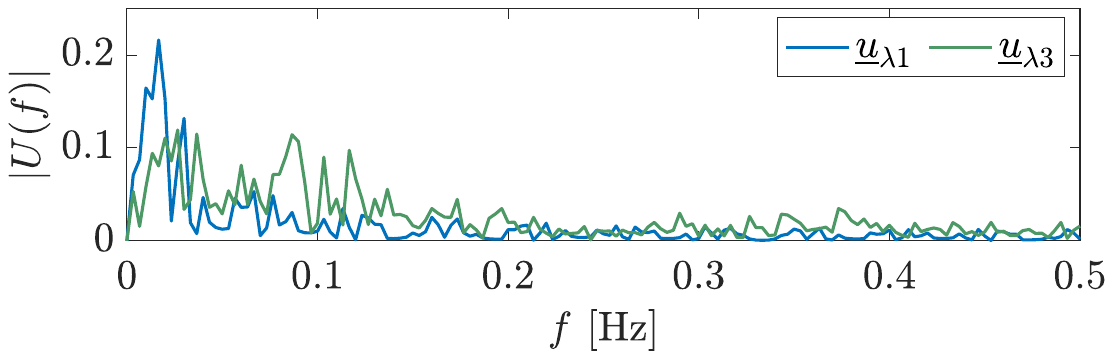}
	\caption{Excited frequency spectrum of $\underline{u}_{\lambda_{1}}$ and $\underline{u}_{\lambda_{3}}$.}
	\label{fig:Spectrum}
\end{figure}

\begin{figure}[b]
	\centering
	\begin{subfigure}[t]{0.2\textwidth}
		\centering
		\includegraphics[width=5cm, trim=3.5cm 7cm 3cm 6.3cm]{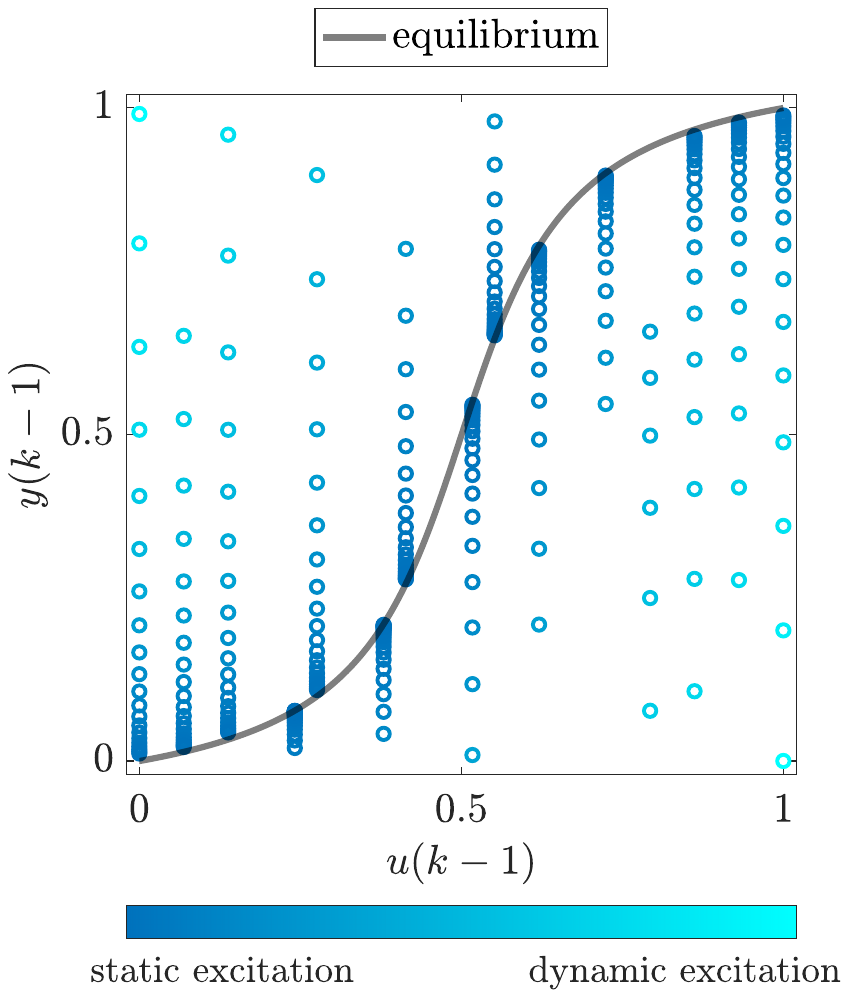}
		\caption{}
		\label{subfig:RspaceLambda0}
	\end{subfigure}
	\hfill 
	\begin{subfigure}[t]{0.2\textwidth}
		\centering
		\includegraphics[width=5cm, trim=3.5cm 7cm 3cm 6.3cm]{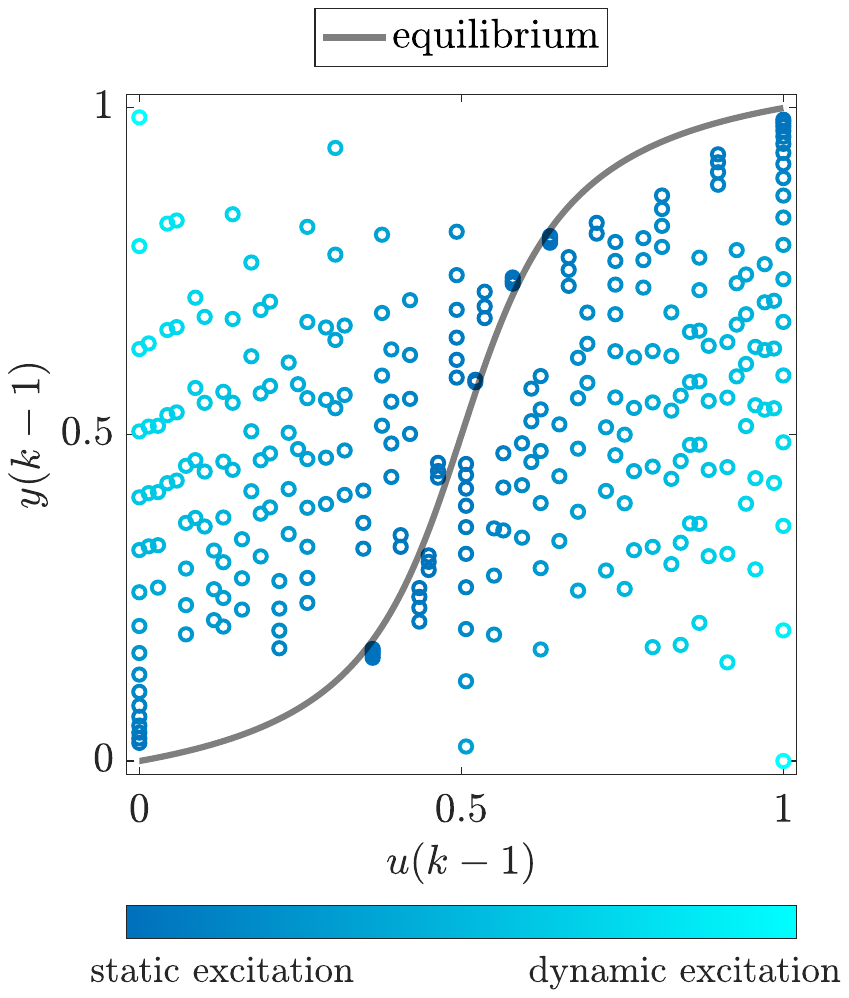}
		\caption{}
		\label{subfig:RspaceLambda05}
	\end{subfigure}
			\caption{Regressor space distribution of $\underline{u}_{\lambda_{1}}$ in (a) and $\underline{u}_{\lambda_{3}}$ in (b).}
	\label{fig:Rspace}
	\end{figure}

The intended purpose of the length-dependent penalty is evident in Fig.$\ $\ref{fig:Signal}, as an increase in $\lambda$ results in the selection of smaller sequence lengths. More precisely, it can be determined that with a growing signal length, which corresponds to the filling of the proxy regressor space, the influence of $J_{2}$ rises and the sequences become increasingly shortened. 
 This change in the trajectory of the excitation signal leads to less frequent dwelling of $y$ on the steady state and manifests more aggressive, dynamic behavior. \\ 
Moreover, the excited spectrum is influenced, as depicted in Fig.$\ $\ref{fig:Spectrum}. 	
From an excitation dominated by low frequencies, it shifts to higher frequencies, i.e.,$\ $to a more dynamic excitation.
Observing the regressor space of the nonlinear approximator in Fig.$\ $\ref*{fig:Rspace}, this is reflected in fewer data points being placed near the equilibrium, but more points in rather dynamic areas.
Apart from the consistent shortening of the sequence lengths for higher $\lambda$-values, the effects of $F$ can be observed. Here, it is important to note that sequences exhibiting a substantial change in their $u$-value tend to exhibit higher dynamics (cf.,$\ $Fig.$\ $\ref*{subfig:RspaceLambda05}). In consideration of this, Fig.$\ $\ref{fig:Signal} (b) reveals that, under the influence of $F$, more dynamic sequences (with significant $u$-changes)  are assigned relatively larger lengths, whereas less dynamic ones (with small $u$-changes) are assigned relatively smaller lengths. As a result, dwelling on the static, contributing limited additional information about the (dynamic) process behavior, is mitigated. Instead, dynamic segments of the sequences are prioritized (cf.$\ $Fig.$\ $\ref{subfig:RspaceLambda05}). Furthermore, scaling with $F$ favors dynamic sequences and for monotone static nonlinearities this results in a preference for large steps (significant $u$-changes), as evident from Fig.$\ $\ref{fig:Signal} (b). \\
As can be seen in Fig.$\  $\ref{fig:ModelOutput}, the different characteristics of the investigated IDS-FID signals influence the model performance. The precise impact on model accuracy is contingent upon the frequency range of process operation:
\begin{figure}[t]
	\begin{center}
		\includegraphics[width=8.4cm, trim=1.7cm 8cm 2.5cm 8cm]{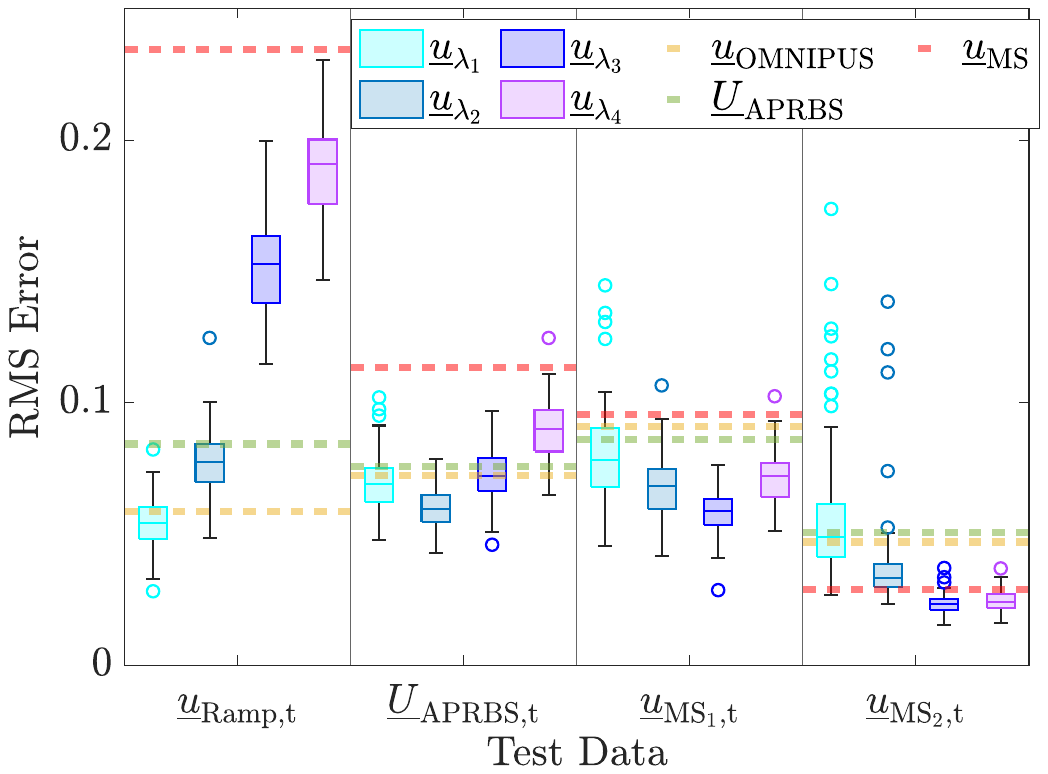}    
		\caption{Performance of LMNs trained with different excitation signals on various test data sets. For $\underline{u}_{\mathrm{OMNIPUS}}, \ \underline{U}_{\mathrm{APRBS}}$, and $\underline{u}_{\mathrm{MS}}$, solely median values are displayed, whereas the RMSEs of all 100 models are presented for $ \underline{u}_{\lambda_{1}}, \ \underline{u}_{\lambda_{2}}, \underline{u}_{\lambda_{3}}$, and $ \underline{u}_{\lambda_{4}}$.}
		\label{fig:ModelOutput}
	\end{center}
\end{figure}
While an increasing $\lambda$ unconditionally improves the model accuracy on $\underline{u}_{\mathrm{MS_{2}, t}}$,  after an initial improvement, the performance deteriorates on $\underline{U}_{\mathrm{APRBS,t}}$ and  $\underline{u}_{\mathrm{MS_{1}, t}}$, due to their lower frequency characteristics.  In parallel, an intensified excitation of higher frequencies worsens the approximation of the steady state, as is evident from the development of the RMSEs on $\underline{u}_{\mathrm{Ramp,t}}$.\\
Hence, it can be established that a more dynamic excitation enhances dynamic model accuracy at the expense of static model accuracy. The excitation should therefore be tailored to the anticipated operating frequency range of the corresponding process, or, in case of unfamiliarity with this information, to the user's model preferences. \\
The comparison to state-of-the-art excitation signals reveals that the IDS-FID strategy generates competitive results even without detailed tailoring to the dynamics of the test data. Fine-tuning of $\lambda$ enhances the approach and it is capable of providing the best excitation on all test data. Solely when a highly dynamic excitation is generated, a significant decrease in approximation quality of the steady state is evident, resulting in the OMNIPUS and the APRBS to clearly outclass the IDS-FID signals generated with $\lambda = [0.5, 2]$ on $\underline{u}_{\mathrm{Ramp,t}}$. Nevertheless, even in this case, the IDS-FID approach demonstrates superior performance compared to the multisine signal. This is remarkable because the multisine signal is the only signal capable of producing results that are similar, albeit inferior, to the IDS-FID signals generated with $\lambda = [0.5, 2]$ in highly dynamic operating areas. The consistently competitive performance across various test data can be attributed to the fulfillment of the space-filling criterion, which ensures that information is collected in all operating regions, leaving no significant knowledge gap about any part of the process. \\
It is worth noting that experiments conducted with varying numbers of data points $N$, different noise characteristics, and employing distinct nonlinear static approximators produced comparable results. Thus, the presented performance evaluation is representative for the outcomes in other investigations. \\
Moreover, it can be affirmed that the IDS-FID algorithm exhibits robustness concerning its parameters $M$ and $n$, i.e.,$\ $(minor) adjustments in their values do not significantly diminish the quality of the emerging excitation signal.

\section{Conclusion and Outlook} \label{SecConclusion}
The evaluation has revealed that tailoring an excitation to match the anticipated frequency range of process operation contributes to model improvement. The IDS-FID strategy has shown high competitiveness compared to other state-of-the-art Design of Experiments (DoE) methods. In future work, the simple linear proxy model is intended to be replaced by more advanced models. Specifically, an online DoE \citep{deflorian2011design} approach will be developed, wherein the proxy model undergoes continuous refinement, taking into account the information gained through signal generation.

\bibliography{ifacconf}             

\end{document}